\documentclass{article}

\usepackage{PRIMEarxiv}

\usepackage[utf8]{inputenc} % allow utf-8 input
\usepackage[T1]{fontenc}    % use 8-bit T1 fonts
\usepackage{hyperref}       % hyperlinks
\usepackage{url}            % simple URL typesetting
\usepackage{booktabs}       % professional-quality tables
\usepackage{amsfonts}       % blackboard math symbols
\usepackage{nicefrac}       % compact symbols for 1/2, etc.
\usepackage{microtype}      % microtypography
\usepackage{lipsum}
\usepackage{fancyhdr}       % header
\usepackage{graphicx}       % graphics
\usepackage{amsmath}
\usepackage{amssymb}
\usepackage{amsthm}
\newtheorem{theorem}{Theorem}

\title{Identification in Endogenous Sequential Treatment Regimes
\thanks{ I thank Cristine Pinto, Jonathan Roth, Peter Hull, Toru Kitagawa, Vitor Possebom, as well as participants in the Brown Econometrics Seminar and the Insper Students Seminar.}}

\author{
  Pedro Picchetti \\
  INSPER \\
  \texttt{pedrop3@al.insper.edu.br} \\
}

\begin{document}
\maketitle

\begin{abstract}
This paper develops a novel nonparametric identification method for treatment effects in settings where individuals self-select into treatment sequences. I propose an identification strategy which relies on a dynamic version of standard Instrumental Variables (IV) assumptions and builds on a dynamic version of the Marginal Treatment Effects (MTE) as the fundamental building block for treatment effects. The main contribution of the paper is to relax assumptions on the support of the observed variables and on unobservable gains of treatment that are present in the dynamic treatment effects literature. Monte Carlo simulation studies illustrate the desirable finite-sample performance of a sieve estimator for MTEs and Average Treatment Effects (ATEs) on a close-to-application simulation study.
\end{abstract}

% keywords can be removed

\section{Introduction}

Evaluating treatment effects is one of the central tasks of modern econometrics. Challenges to the identification of treatment effects come from the fact that individuals that decide to take treatment are generally different from the ones that do not enter treatment. Econometric causal inference methods try to deal with the identification challenge that comes from self-selection into treatment by exploiting variation in treatment status that is independent from the gains of treatment: Instrumental variables, parallel trends, discontinuities in treatment assignment are some of the most pervasive approaches used by applied economists.

Most identification strategies focus on settings with cross-sectional data and a binary treatment. However, identification in settings where panel data is available and agents to choose to enter treatment or not multiple times is far more challenging. Examples of such settings are multi-period clinical trials, multi-period academic interventions and alternative sentence regimes for repeated criminal offenders. The dynamic nature of choice and outcomes impose additional challenges to identification: agents endogenously make decisions of treatment at each period with at least partial knowledge of the expected gains, while learning about the gains from past outcomes and treatment choices. Treatment effects might be heterogeneous with respect to the amount of times agents received treatment (dose-response), at the same time that there can be heterogeneity associated to taking treatment in a particular period (calendar effects).

This paper proposes a novel nonparametric identification method to identify the effects of sequences of treatments using dynamic instrumental variables (IV). The method identifies a dynamic version of the marginal treatment effect (MTE) that serves as a building block for the average treatment effect (ATE) associated to a sequence of treatments. To put it simply, the method combines the intuition of the cross-sectional MTE approach  (\cite{heckman2001};\cite{heckman2005}) to deal with the endogeneity that arises from self-selection with the g-formula method (\cite{ROBINS19861393}).

Dynamic treatment effects in sequential treatment regimes have been widely studied in the field of biostatistics (\cite{ROBINS19861393};\cite{ROBINS1987139S};\cite{murphy};\cite{hernan}). Identification strategies in this literature rely on the sequential randomization assumption, a dynamic version of random assignment with perfect compliance. The sequential randomization assumption is unlikely to hold in most economic settings. Non-compliance can be prevalent due to the fact that compliance at every period can be costly for agents and that agents can learn about the gains of treatment and thus self-select into treatment.

In the Econometrics field, sequential treatment regimes have received less attention. In the Difference-in-Differences literature,\cite{frenchguys2} and \cite{DECHAISEMARTIN2023105480} propose an identification strategy that combines a strong exogeneity assumption (potential outcomes being independent from treatment sequence) with the parallel trends assumption. The approach, however, is not valid in the presence of selection based on expected gains, which violates the strong exogeneity assumption. Moreover, even if one assumes that the data satisfies these high-level assumptions, dynamic choice models that allow for entering and exiting treatment are not compatible with the standard parallel trends assumption (\cite{ghanem2023selection}; \cite{marx2023parallel}). Thus, the method lacks solid foundation from a choice-theoretic perspective.

In the instrumental variables approach, \cite{HAN2021132} provides an identification method for the average treatment effect observed at the last period available in the sample. The identification strategy is a multi-period version from the identification strategy from \cite{yildiz} that relies on sequential rank similarity conditions and additional exclusion restrictions or further assumptions on the support of the covariates. Thus, the method rules out the possibility of unobservable gains and requires a covariate in the potential outcomes model that is excluded from the choice model, or sufficient variation in the covariates to identify the ATE.

In this paper, I relax the rank similarity and the support assumptions by focusing on the identification of a more local treatment effect, which is the MTE. I show that the MTE can be identified under the dynamic version of the exclusion restriction and under sufficient variation of the instrument or additional functional form assumptions, it can be used as a building block for the ATE, which is obtained by integrating the MTE.

First, I show that is not possible to identify MTEs in sequential treatment regimes modifying the local instrumental variable (LIV) estimand. That occurs because the number of potential outcomes grows exponentially with the number of periods in the sample. The identification approach builds on the marginal treatment response (MTR) functions, which generate MTEs (see \cite{mogstad} for a discussion on MTRs in cross-sectional data). The MTRs are identified conditional on time-varying confounders (eg: intermediary outcomes), and I show that the time-varying confounders can be eliminated from the MTRs by using a version of the g-formula.

Monte Carlo simulation studies assert the desirable finite-sample performance of a sieve estimator for dynamic MTEs and ATEs in a simulated DGP that simulates a two-period setting in which the dynamic MTEs and agents learn about the expected gains of treatment.

\textbf{Organization of the paper:} In the next section, I present the main framework of the paper. Section 3 presents the main identification results, followed by a discussion of motivating examples in Section 4. Section 5 presents the Monte Carlo simulation studies and Section 6 concludes.

\section{Framework}

Before I introduce the framework, some notation is required. Panel data is available for $\mathcal{T}$ periods, with a particular time period being denoted by $t=1,...,\mathcal{T}$. For a random variable $R$, we denote $R_{t}$ as the observation of R at period t and $r_{t}$ as its realization. Let $\textbf{R}^{t}=(R_{1},...,R_{t})$ denote te row vector that collects $R_{t}$ until time t, and $\textbf{r}^{t}$ denote its realization. Furthermore, let $\textbf{R}=\textbf{R}^{\mathcal{T}}$ for convenience.

Consider the following dynamic structural model for outcomes in period t, which is a function of the sequence of past outcomes $\textbf{Y}^{t-1}$, the sequence of treatments until period t, $\textbf{D}^{t}$, baseline covariates $X$ and an error term that is correlated to the sequence of treatments $U_{t}(\textbf{D}^{t})$:

\begin{equation}
    Y_{t}=\mu_{t}(\textbf{Y}^{t-1},\textbf{D}^{t},X,U_{t}(\textbf{D}^{t}))
\end{equation}

Implicit in equation 1 lies a no-anticipation assumption, which is formalized below

\textbf{Assumption NA:} For $t=1,...,\mathcal{T}$, $Y_{t}=\sum_{\textbf{d}^{t}\in\left \{ 0,1 \right \}^t}\mathbf{1}\left \{ \textbf{D}^{t}=\textbf{d}^{t} \right \}Y_{t}(\textbf{d}^{t})$

Assumption \textbf{NA} states that the potential outcomes in period $t$ are a function of treatment sequences until $t$, which is a pervasive assumption in the dynamic treatment effects literature. Note that the definition of potential outcomes as a function of the sequence of treatments is flexible enough so that the timing of treatment might play a role in determining the outcomes, as the same time as it can simplify to a setting in which the amount of times the agent receives treatment captures the whole heterogeneity in potential outcomes, thus collapsing to a dose-response framework.

Correlation between the unobserved term $U_{t}$ and the sequence of treatments comes from the fact that individuals choose treatment at each period with at least partial knowledge of the expected gains from treatment. I introduce the following single-threshold crossing model for selection into treatment at each period, which is a function of past outcomes, the past sequence of treatments, the sequence of instruments until period t and baseline covariates:

\begin{equation}
    D_{t}=\mathbf{1}\left \{ \Psi_{t}(\textbf{Y}^{t-1},\textbf{D}^{t-1},\textbf{Z}^{t},X)\geq \eta_{t} \right \}
\end{equation}

Correlation between $U_{t}$ and $\textbf{D}^{t}$ arises because $U_{t}$ is correlated to the vector of unobservables driving selection into treatment until that period, denoted by $\mathbf{\eta}^{t}$. Unlike \cite{HAN2021132}, I do not restrict the distribution of $U_{t}(\textbf{d}^{t})$ and $U_{t}(\textbf{d}^{t'})$ to be the same for $\textbf{d}^{t}\neq \textbf{d'}^{t}$ conditional on $\mathbf{\eta}^{t}$.

Below, we invoke a regularity assumption on the vector $\mathcal{\eta}$ that enables the standard normalization of the period-specific choice model.

\textbf{Assumption C:} For $t=1,...,\mathcal{T}$, $\eta_{t}$ is absolutely continuous with respect to the Lebesgue measure. 

Under assumption $\textbf{C}$, it is possible to write the choice model as 

\begin{equation}
    D_{t}=\mathbf{1}\left \{ \pi_{t}(\textbf{Y}^{t-1},\textbf{D}^{t-1},\textbf{Z}^{t},X)\geq V_{t} \right \}
\end{equation}

Where $V_{t}:= F_{\eta_{t}|\textbf{Y}^{t-1}\textbf{D}^{t-1},\textbf{Z}^{t-1},X}(\eta_{t})$ has a standard uniform distribution and can be interpreted as the quantile of resistance to treatment at period t, while $\pi_{t}(\textbf{Y}^{t-1}\textbf{D}^{t-1},\textbf{Z}^{t},X):= F_{\eta_{t}|\textbf{Y}^{t-1}\textbf{D}^{t-1},\textbf{Z}^{t-1},X}(\Psi_{t}(\textbf{Y}^{t-1}\textbf{D}^{t-1},\textbf{Z}^{t},X))$ is also bounded within the unit interval and has a propensity score interpretation. This sort of choice model imply monotonicity with respect to $Z_{t}$ conditional on the observed data from previous periods and baseline covariates (\cite{han23}).

The main parameter of interest in this paper is a version of the marginal treatment effect (MTE), which I define as

\begin{equation}
    MTE(\textbf{d}^{t},\textbf{d`}^{t};x,\textbf{v}^{t})=\mathbb{E}\left [ Y_{t}(\textbf{d}^{t})-Y_{t}(\textbf{d`}^{t})|X=x,\textbf{V}^{t}=\textbf{v}^{t} \right ]
\end{equation}

which is the average difference between potential outcomes of two different treatment regimes conditional on the baseline covariates and the vector of unobservables that drive selection. Much like in the cross-sectional framework, the dynamic MTE is the causal parameter that connects the potential outcomes model with the choice model. 

The MTE holds several desirable properties. It is a function that is invariant to the choice of the instrument and holds a straightforward economic interpretation as the willingness to pay for treatment. With cross-sectional data, evaluating the MTE over the range of the distribution of $V$ allows for the full recovery of heterogeneity in treatment effects across the population and provides insights on the pattern of selection on gains.

With panel data and sequential treatment regimes, the recovery of the heterogeneity in treatment effects provides not only insights on the pattern of selection on gains, but also on how they change with time and different treatment sequences. For example, it is possible to identify if agents are learning about the expected gains of treatment with time.

Furthermore, the MTE is also a building block for the other average treatment effects in the causal inference literature, as they can be expressed as weighted averages of the MTE. In sequential treatment regimes, the MTE hold the same property. In this, paper I focus on the identification of dynamic ATEs using dynamic MTEs as the building blocks. ATEs in this setting are expressed as

\begin{equation}
    ATE(\textbf{d}^{t},\textbf{d`}^{t};x):= \mathbb{E}\left [ Y_{t}(\textbf{d}^{t})-Y_{t}(\textbf{d`}^{t})|X=x \right ]=\int_{\textbf{v}^{t}}MTE(\textbf{d}^{t},\textbf{d`}^{t};x,\textbf{v}^{t})d\textbf{v}^{t}
\end{equation}

which is nonparametrically identified if the period-specific instruments are continuous or under additional functional form assumptions under a dynamic version of the exclusion restriction, formalized below.

\textbf{Assumption SX:} $(\textbf{U}(\textbf{d}),\textbf{V})\perp\textbf{Z}|X$ for all $\textbf{d}\in\left \{ 0,1 \right \}^{\mathcal{T}}$. 

Assumption $\textbf{SX}$ is dynamic version of the conditional independence assumption that appears in the cross-sectional MTE framework. The assumption is satisfied in multi-period experiments randomized trials and multi-period quasi-experiments. A canonical example would be the sequence of treatment assignments in a multi-period clinical trial.

An alternative specification for potential outcomes in sequential treatment regimes would be through a structural function of covariates and an error term that varies across sequences of treatments:

\begin{equation*}
    Y_{t}(\textbf{d}^{t})=\mu_{\textbf{d}^{t}}(X,U_{t}(\textbf{d}^{t})), \textbf{d}^{t}\in\left \{ 0,1 \right \}^{t}
\end{equation*}

Such an specification is more flexible in that sense that potential outcomes do not depend on previous outcomes. A compatible choice model for treatment at each period would be

\begin{equation*}
    D_{t}=\mathbf{1}\left \{ \pi_{t}(\textbf{D}^{t-1},\textbf{Z}^{t},X)\geq V_{t} \right \}
\end{equation*}

Under this specification, it is implied that individuals do not learn from their previous outcomes, but still learn from their previous treatment choices. In the next section, I present identification results suited for the recursive structural models in which present outcomes depend on the history of outcomes, but the results can be easily modified to identify treatment effects in the latter specification of potential outcomes as well.

\section{Identification}

Identification in this setting uses the dynamic marginal treatment response (MTR) function as the fundamental parameter. The dynamic MTR is defined as

\begin{equation}
    m_{\textbf{d}^{t}}(x,\textbf{v}^{t}):=\mathbb{E}\left [ Y_{t}(\textbf{d}^{t})|X=x,\textbf{V}^{t}={\textbf{v}}^{t} \right ]
\end{equation}

and a pair of MTRs generates a MTE: $MTE(\textbf{d}^{t},\textbf{d'}^{t};x,\textbf{v}^{t})=m_{\textbf{d}^{t}}(x,\textbf{v}^{t})-m_{\textbf{d'}^{t}}(x,\textbf{v}^{t})$. In the cross-sectional MTE framework, MTEs can be identifide using the local instrumental variable (LIV) estimand (\cite{heckman2001}) or through the identification of the MTRs separately (\cite{brinch}; \cite{mogstad}). In sequential treamnet regimes, however, a dynamic version of LIV does not recover the MTEs.

I illustrate the no-identification under the LIV estimand and the identification of MTRs in a simple two-period example. From this point forward, covariates are supressed from notation for the sake of simplicity, but all results hold conditional on X.

\subsection{Two-period toy model}

Consider the case where panel data is available for two-periods. Individuals choose to enter treatment or not at each period, which means that in the second period there are four potential outcomes:$Y_{2}(0,0), Y_{2}(1,0), Y_{2}(0,1), Y_{2}(1,1)$. The firs potential outcome is associated to not entering into treatment in any period, the second one to exposure to treatment in the first period, the third to exposure in the second period and the last is associated to full exposure to treatment. The observed outcome in period 2 is

\begin{align*}
    Y_{2}=D_{1}D_{2}Y_{2}(1,1)+(1-D1)D_{2}Y_{2}(0,1)+D_{1}(1-D_{2})Y_{2}(1,0)+(1-D1)(1-D_{2})Y_{2}(0,0)\\=Y_{2}(0,0)+D_{1}D_{2}(Y_{2}(1,1)+Y_{2}(0,0)-Y_{2}(0,1)-Y_{2}(1,0))+D_{1}(Y_{2}(1,0)-Y_{2}(0,0))+D_{2}(Y_{2}(0,1)-Y_{2}(0,0))
\end{align*}

Under Assumptions $\textbf{C}$ and $\textbf{SX}$, the period specific propensity scores $\pi_{1}(Z_{1})$ and $\pi_{2}(Y_{1},D_{1},\textbf{Z}^{2})$ are identified from the data. Under Assumptions $\textbf{NA}$, $\textbf{C}$ and $\textbf{SX}$ a dynamic version of the LIV estimand, that is, $\frac{\partial \mathbb{E}\left [ Y_{2}|Y_{1},\pi_{1}(Z_{1}),\pi_{2}(Y_{1},D_{1},\textbf{Z}^{2}) \right ]}{\partial \pi_{1}(Z_{1})\partial \pi_{2}(Y_{1},D_{1},\textbf{Z}^{2})}$, identifies the following quantity

\begin{align*}
    \mathbb{E}\left [Y_{2}(1,1)+Y_{2}(0,0)-Y_{2}(1,0)-Y_{2}(0,1)|Y_{1},\textbf{V}^{2}=(\pi_{1}(Z_{1}),\pi_{2}(Y_{1},D_{1},\textbf{Z}^{2})) \right ]\\+\frac{\partial \mathbb{E}\left [ Y_{2}(1,0)-Y_{2}(0,0)|Y_{1},V_{1}=\pi_{1}(Z_{1}),\pi_{2}(Y_{1},D_{1},\textbf{Z}^{2}) \right ]}{\partial \pi_{2}(Y_{1},D_{1},\textbf{Z}^{2})}\\+\frac{\partial \mathbb{E}\left [ Y_{2}(1,0)-Y_{2}(0,0)|Y_{1},\pi_{1}(Z_{1}),V_{2}=\pi_{2}(Y_{1},D_{1},\textbf{Z}^{2}) \right ]}{\partial \pi_{1}(Z_{1})}
\end{align*}

which holds no clear causal interpretation. See the appendix for a proof of the result. While identification of dynamic MTEs with a dynamic version of the LIV estimand is not possible in sequential treatment regimes, identification via generating two MTRs is straightforward. To illustrate the procedure, suppose we are interested in the effect of being exposed to treatment in the second period versus the first period:$MTE((0,1),(1,0);\textbf{V}^{2}=\textbf{v}^{2})$ and $ATE((0,1),(1,0))$.

These treatment effects are interesting because they can provide insights on the heterogeneity that comes from the timing of treatment and on how agents learn about the expected gain from treatment from period 1 to period 2. Identification of the MTRs follows a dynamic version of the identification approach in \cite{brinch}. For the MTR associated to treatment in the second period, we have

\begin{equation*}
    -\frac{\partial \mathbb{E}\left [ (1-D_{1})D_{2}Y_{2}|Y_{1},\pi_{1}(Z_{1}),\pi_{2}(Y_{1},D_{1},\textbf{Z}^{2}) \right ]}{\partial \pi_{1}(Z_{1}),\pi_{2}(Y_{1},D_{1},\textbf{Z}^{2})]}=\mathbb{E}\left [ Y_{2}(0,1)|Y_{1},\textbf{V}^{2}=(\pi_{1}(Z_{1}),\pi_{2}(Y_{1},D_{1},\textbf{Z}^{2})) \right ]
\end{equation*}

Which is the MTR associated to that sequence of treatment conditional on the intermediary outcome $Y_{1}$. To obtain the MTR of interest, one can interpret $Y_{1}$ as a time-varying confounder and apply the g-formula to the result above 

\begin{align*}
    -\int_{y_{1}}\frac{\partial \mathbb{E}\left [ (1-D_{1})D_{2}Y_{2}|Y_{1},\pi_{1}(Z_{1}),\pi_{2}(Y_{1},D_{1},\textbf{Z}^{2}) \right ]}{\partial \pi_{1}(Z_{1}),\pi_{2}(Y_{1},D_{1},\textbf{Z}^{2})]}dF_{Y_{1}|D_{1},\pi_{1}(Z_{1})}(y_{1}|d_{1},\pi_{1}(z_{1}))\\=\mathbb{E}\left [ Y_{2}(0,1)|\textbf{V}^{2}=(\pi_{1}(Z_{1}),\pi_{2}(Y_{1},D_{1},\textbf{Z}^{2})) \right ]=m_{(0,1)}(\pi_{1}(Z_{1}),\pi_{2}(Y_{1},D_{1},\textbf{Z}^2))
\end{align*}

Similarly, we can identify $m_{(1,0)}(\pi_{1}(Z_{1}),\pi_{2}(Y_{1},D_{1},\textbf{Z}^{2}))$ and generate $MTE((0,1),(1,0);\pi_{1}(Z_{1}),\pi_{2}(Y_{1},D_{1},\textbf{Z}^{2}))$. The ATE is identified by integrating the MTRs over the range of the vector of unobservables $\textbf{V}^{2}$:

\begin{equation*}
    ATE((0,1),(1,0))=\int_{0}^{1}\int_{0}^{1}m_{(0,1)}(\textbf{v}^{2})dv_{1}dv_{2}-\int_{0}^{1}\int_{0}^{1}m_{(1,0)}(\textbf{v}^{2})dv_{1}dv_{2}
\end{equation*}

If the instruments are continuous, then the whole support of the propensity score is identified, so the MTEs and consequently the ATEs can be identified without any further assumptions. If the instruments are discrete, however, further assumptions are required. In the next section I formalize the identification results.

\subsection{Main Identification Result}

I now formalize the identification result from the procedure outlined in Section 3.1. I follow \cite{mogstad} and assume the researcher restricted the MTRs $m_{\textbf{d}^{t}}(\textbf{v}^{t})$ to lie in some parameter space  $\mathcal{M}$ that incorporates assumptions about functional forms of potential outcomes. For instance, if instruments are continuous at all periods, the parameter space $\mathcal{M}$ can be left unrestricted. If the instruments are binary, 
 it follows form \cite{brinch} that absent from further assumptions, the parameter space $\mathcal{M}$ only contains MTRs that are linear in the vector $\textbf{v}^{t}$. In order to identify MTRs with more general functional forms, then the researcher must restrict the parameter space $\mathcal{M}$ to contain only MTRs that are additively separable between $X$ and $\textbf{V}^{t}$. Before introducing the main identification result, define $\mathcal{\pi}^{t}(\textbf{Y}^{t-1},\textbf{Z}^{t})$ as the vector of period-specific propensity scores until period t.

 \begin{theorem}
     Suppose assumptions $\textbf{NA}$, $\textbf{C}$ and $\textbf{SX}$ hold. Then for all $t=1,...,\mathcal{T}$,

     \begin{equation*}
         \int_{\textbf{y}^{t-1}}\frac{\partial \mathbb{E}\left [ \mathbf{1}\left \{ \textbf{D}^{t}=\textbf{d}^{t} \right \}Y_{t}|\textbf{Y}^{t-1}\pi^{t}(\textbf{Y}^{t-1},\textbf{D}^{t-1},\textbf{Z}^{t}) \right ]}{\partial \pi_{1}(Z_{1})...\partial\pi_{t}(\textbf{Y}^{t-1},\textbf{D}^{t-1},\textbf{Z}^{t})}\prod_{k=1}^{t-1}dF_{Y_{k}|\textbf{Y}^{k-1},\textbf{D}^{k},\textbf{Z}^{k}}(y_{k}|\textbf{y}^{k-1},\textbf{d}^{k},\textbf{z}^{k}))=m_{\textbf{d}^{t}}
(\pi^{t}(\textbf{Y}^{t-1},\textbf{D}^{t-1},\textbf{Z}^{t}))
     \end{equation*}

     If the number of untreated periods is even, and

     \begin{equation*}
         \int_{\textbf{y}^{t-1}}-\frac{\partial \mathbb{E}\left [ \mathbf{1}\left \{ \textbf{D}^{t}=\textbf{d}^{t} \right \}Y_{t}|\textbf{Y}^{t-1}\pi^{t}(\textbf{Y}^{t-1},\textbf{D}^{t-1},\textbf{Z}^{t}) \right ]}{\partial \pi_{1}(Z_{1})...\partial\pi_{t}(\textbf{Y}^{t-1},\textbf{D}^{t-1},\textbf{Z}^{t})}\prod_{k=1}^{k-1}dF_{Y_{k}|\textbf{Y}^{k-1},\textbf{D}^{k},\textbf{Z}^{k}}(y_{k}|\textbf{y}^{k-1},\textbf{d}^{k},\textbf{z}^{k}))=m_{\textbf{d}^{t}}
(\pi^{t}(\textbf{Y}^{t-1},\textbf{D}^{t-1},\textbf{Z}^{t}))
     \end{equation*}

     If the number of untreated periods is odd, for all $m_{\textbf{d}^{t}}(\textbf{v}^{t})\in\mathcal{M}$.
 \end{theorem}

Theorem 1 provides the main identification result in this paper. It shows that the dynamic MTRs at each period can be identified by differentiating conditional expected outcomes at period t with respect to the vector of propensity scores until period t and by integrating over the support of the vector of intermediary outcomes conditional on the observed data from previous periods. Its thus a procedure that combines the intuition of the cross-sectional framework with the g-formula. The multiplication by minus 1 in the result from an odd number of untreated periods follows from the fact that when applying the Leibinz rule in the differentiation of the integral of the conditional expectation, the derivative equals the negative of the MTR. See the proof in the Appendix for more details.

Note that in settings in which potential outcomes do not depend on the history of past outcomes, there are no time-varying confounders, and thus the MTRs can be directly identified using the differentiation of the vector of propensity scores, and the use of the g-formula is not necessary:

\begin{equation*}
    \frac{\partial \mathbb{E}\left [ \mathbf{1}\left \{ \textbf{D}^{t}=\textbf{d}^{t} \right \}Y_{t}|\pi^{t}(\textbf{Z}^{t},\textbf{D}^{t-1}) \right ]}{\partial \pi_{1}(Z_{1})...\partial \pi_{t}(\textbf{Z}^{t})}=m_{\textbf{d}^{t}}(\pi^{t}(\textbf{Z}^{t},\textbf{D}^{t-1}))
\end{equation*}

 for a even number of untreated periods and

 \begin{equation*}
     -\frac{\partial \mathbb{E}\left [ \mathbf{1}\left \{ \textbf{D}^{t}=\textbf{d}^{t} \right \}Y_{t}|\pi^{t}(\textbf{Z}^{t},\textbf{D}^{t-1}) \right ]}{\partial \pi_{1}(Z_{1})...\partial \pi_{t}(\textbf{Z}^{t})}=m_{\textbf{d}^{t}}(\pi^{t}(\textbf{Z}^{t},\textbf{D}^{t-1}))
 \end{equation*}

 for an odd number of untreated periods.
 
 A priori one specification is not preferred to another, and it is up to the applied researcher to chose the one that suits best its applied setting. The choice will depend mostly on how individuals are assumed to learn dynamically the particular empirical environment. In the next section, I discuss some examples of empirical works in economics under which this identification approach can be employed.

\section{Motivating Examples}

The most straightforward example of a setting of endogenous sequential treatment regime would be a multi-period experiment with imperfect compliance.
Take for instance the Fast Track Program, an extra-curricular teacher consultation program established in the 1990s by the National Institute of Mental Health to prevent drug use in by children at risk. \cite{murphy} measure the dynamic effects of the program using identification methods that rely on the sequential randomization assumption. Thus, the results can be interpreted as dynamic intention-to-treat effects.

Experimental settings aside, sequential treatment regimes are still very pervasive in applied economics.

There is an extensive literature on the economics of education that focuses on measuring the impact of attending certain types of education institutions on educational outcomes. In the US, there is a vast literature estimating the impact of attending charter schools  using multi-period lotteries as instruments (\cite{angrist,curto,dobbie}). In France, \cite{boarding} analyze the impact of boarding schools on educational outcomes also using multi-period lotteries as as instruments. In these papers, students can comply or not to the lottery results, and they can enter and leave treatment. In \cite{boarding}, for example, 24\% of the students left the boarding school at the end of their first year enrolled.

In the literature of the economics of crime, sequential treatments appear in the literature studying the impact of sentences on recidivism and labor market outcomes, which uses the so-called judge fixed effects as instruments for sentence decisions. \cite{magne} measures the impact of sentencing to prison on future criminal behavior and employment. The authors follow a event-study approach by normalizing the period of the court decision to zero. Due to sample size issues the authors refrain from separating first-time criminals from repeated offenders, but one could apply the identification strategy presented in this paper to repeated offenders to measure how the effect of a sequence of court decisions affects labor market outcomes.

These are just some examples in which endogenous sequential treatments appear in the applied economics literature. Whenever there are dynamic instrumental variables available, the identification strategy presented in Section 3.2 can be employed to measure the heterogeneity in treatment effects that comes from the unobservable gains of treatment and the dynamic learning from previous choices, as well as the heterogeneity that comes from the time of treatment choices.

For example, take the charter school examples. It could be that the educational benefits from attending a charter school in a given grade fade if you leave the charter school. There could be a reverse selection on gains in the first period of the lottery, but for the next period individuals learn about the gains from attending the charter school, and thus the pattern of heterogeneity in the unobservable gains changes with time.

In the next section, I illustrate this feature in a Monte Carlo simulation study of a close-to-application data generating process (DGP).

\section{Close-to-Application Monte Carlo Simulations}

In this section, I show the desirable finite-sample properties of a series estimator for dynamic MTEs using a close-to-application simulation of a DGP that resembles the data from the Fast Track Program\footnote{I reached out to the Center for Child and Family Policy from Duke University requesting the data from the program but did not get access until the time of submission.}.

The Fast Track Program was established in 1990 by the National Institute of Mental Health to prevent drug use and misconduct among children at risk. The program consists of a multi-period randomized trial that included 891 children from the US. Children were randomized at each period into a treatment and a control group. The treatment consists in a multi-level intervention directed to the children (emotional skills training and academic tutoring), their parents (management training) and teachers (classroom management training and a curriculum for misconduct prevention).

Taking the treatment can be costly in terms of time and effort, and thus imperfect compliance is a big issue in the program. As shown by \cite{murphy}, about 50\% of the households assigned to treatment did not take part did not follow the intervention.

The program follows individuals from the age of 5 until they are 25 years old. For the sake of simplicity, I simulate data from the program for two periods. The simulated data should be interpreted as data for the two first years of elementary school. The simulated outcome is binary and stands for an indicator for school behavior problems. I simulate two binary covariates that represent indicators of race and suspected abuse at kindergarten capturing background characteristics, and also a continuous indicator ranging from 0 to 10 that represents a weighted average of the children`s grades, capturing thus academic performance. Treatment at each period is an indicator for whether children took part in the program at that period, and the binary instrument at each period is the randomized assignment from the program. I set the ATE to be equal to zero for all sequences of treatment, and the MTEs to be linear. The MTE is a upward-sloping function of the unobservable from the choice model in period 1, which indicates a pattern of reverse selection on gains in the first period of treatment, whereas it is a downward-sloping function of the unobservable from the second period choice model, which indicates positive selection on gains in period 2.

For this design, we consider 10.000 Monte Carlo Simulation studies. The simulated DGPs follow the sequential structural function from Section 2.2. As an initial exercise, I study the properties of the MTE associated with taking the program in both periods vs not taking the program at all: $\mathbb{E}\left [ Y_{2}(1,1)-Y_{2}(0,0)|X=x,\textbf{V}^{2}=\textbf{v}^{2} \right ]$.

In Figure 1, I fix a value for $V_{2}$ and the covariates and show the curve of the MTE as a function of $V_{1}$. 

\begin{figure}[htb]
  \centering
  \caption{MTE curve - Unobservables driving the first choice}
  \includegraphics[scale=0.95]{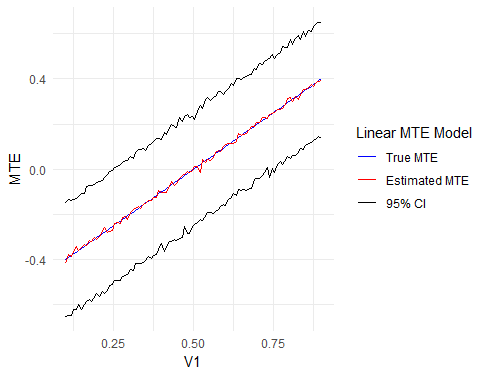}
  \scriptsize \noindent\\ \textit{Note:} Figure 3 depicts the MTE curve for returns to primary education estimated by the parametric polynomial method. The 95\% confidence interval is based on nonparametric bootstrapped standard errors.
\end{figure}

The power series estimator successfully recovers the upward-sloping pattern of the MTE function with respect to $V_{1}$. Confidence intervals were obtained through 999 bootstrap replications. The estimator appears to consistently recover MTE estimates for all values of the propensity score. Confidence intervals are slightly greater at extreme values of the propensity score, which is likely due to the fact the common support loses its strength at such extreme values. 

\begin{figure}[htb]
  \centering
  \caption{MTE curve - unobservables driving the second choice}
  \includegraphics[scale=0.95]{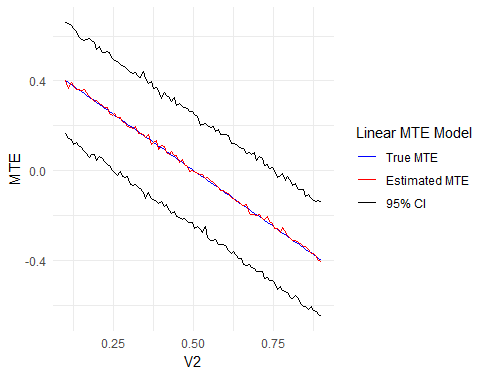}
  \scriptsize \noindent\\ \textit{Note:} Figure 3 depicts the MTE curve for returns to primary education estimated by the parametric polynomial method. The 95\% confidence interval is based on nonparametric bootstrapped standard errors.
\end{figure}

In Figure 2, I fix a value for $V_{1}$ and the covariates and show the curve of the MTE as a function of $V_{2}$. Results are similar to the ones presented in Figure 1. The estimator recovers the sign and magnitude of the MTE at all values of the $V_{2}$ distribution. Confidence intervals are slightly wider.

In Table 1 I present the results of the Monte Carlo simulations for the MTEs for different sequences of treatments evaluated at different points of the vector of resistances to treatment at each period.

\begin{table}[h]
\centering
\caption{Average bias for different values of the MTE function}
\begin{tabular}{c|ccc}
\hline
MTE Functions                & $v^{2}=(0.5,0.5)$ & $v^{2}=(0.25,0.75)$ & $v^{2}=(0.75,0.25)$ \\ \hline
MTE((1,1),(0,0);$V^{2}=v^{2}$) & 0.005           & -0.001            & 0.002             \\
MTE((1,0),(0,0);$V^{2}=v^{2}$) & 0.002           & -0.007            & 0.008             \\
MTE((0,1),(0,0);$V^{2}=v^{2}$) & 0.006           & 0.004             & -0.007            \\
MTE((1,0),(0,1);$V^{2}=v^{2}$) & -0.001          & -0.008            & -0.001            \\ \hline
\end{tabular}
\\
\scriptsize \noindent \textit{Note:} Simulations based on 10.000 Monte Carlo experiments with sample size $n=891$ and $T=2$. Estimates were obtained using a sieve estimator. MTEs are evaluated at the mean value of the covariates from the DGP.
\end{table}

The estimator`s performance analyzed through its average bias is roughly constant across sequences of treatment and different values of the vector of distastes, which shows that it successfully estimates dynamic MTEs for different sequences and is thus able to recover the full heterogeneity in treatment effects associated to this setting.

The dynamic MTEs are a building block for the dynamic ATEs. In Table 2, I show that the estimates for dynamic ATEs obtained by integrating the dynamic MTEs exhibit desirable finite-sample properties. 

\begin{table}[h]
\centering
\caption{Monte Carlo Simulation Result for ATEs}
\begin{tabular}{c|ccccc}
\hline
ATEs             & Av. Bias & Med. Bias & RMSE   & Cover  & CIL    \\ \hline
ATE((1,1),(0,0)) & -0.001  & -0.001   & 0.029 & 0.938 & 0.358 \\
ATE((1,0),(0,0)) & 0.005   & -0.003   & 0.033 & 0.928 & 0.371 \\
ATE((0,1),(0,0)) & -0.002  & -0.003   & 0.011 & 0.956  & 0.312 \\
ATE((1,0),(0,1)) & -0.006  & 0.004    & 0.031 & 0.952  & 0.385 \\ \hline
\end{tabular}\\
\scriptsize \noindent \textit{Note:} Simulations based on 10,000 Monte Carlo experiments. “Av. Bias”, “Med. Bias”, “RMSE”, “Cover” and “CIL’, stand for the average simulated bias, median simulated bias, simulated
root mean-squared errors, 95\% coverage probability, and 95\% confidence interval length,
respectively.
\end{table}

The average bias, median bias and root mean-squared error for the ATEs are small, and the coverage of the boostrapped confidence interval is close to the desired 95\%. Confidence interval lenghts are fairly stable across the considered sequences of treatments.

In table 3 I present the results from Monte Carlo simulations using the parametric g-formula from Robins (1987), implemented using the 'gfoRmula' package designed for the R software. Simulations are biased for all sequences of treatment considered, which was expected as the sequential randomization assumption that is fundamental for the validity of the standard g-formula does not hold in our setting. Confidence intervals are tighter, but empirical coverage is far from the desired 95\%.

\begin{table}[h]
\centering
\caption{Monte Carlo Simulation Result for ATEs - g-formula}
\begin{tabular}{c|ccccc}
\hline
ATEs             & Av. Bias & Med. Bias & RMSE   & Cover  & CIL    \\ \hline
ATE((1,1),(0,0)) & 0.431  & 0.567   & 0.745 & 0.452 & 0.284 \\
ATE((1,0),(0,0)) & -0.555   & -0.459   & 0.863 & 0.567 & 0.277 \\
ATE((0,1),(0,0)) & -0.410  & -0.395   & 0.670 & 0.438  & 0.303 \\
ATE((1,0),(0,1)) & 0.503  & 0.654    & 0.812 & 0.511  & 0.283 \\ \hline
\end{tabular}\\
\scriptsize \noindent \textit{Note:} Simulations based on 10,000 Monte Carlo experiments. “Av. Bias”, “Med. Bias”, “RMSE”, “Cover” and “CIL’, stand for the average simulated bias, median simulated bias, simulated
root mean-squared errors, 95\% coverage probability, and 95\% confidence interval length,
respectively.
\end{table}

\begin{table}[h]
\centering
\caption{Monte Carlo Simulation Result for ATEs - Dynamic DiD}
\begin{tabular}{c|ccccc}
\hline
ATEs             & Av. Bias & Med. Bias & RMSE   & Cover  & CIL    \\ \hline
ATE((1,1),(0,0)) & 0.455  & 0.503   & 0.749 & 0.783 & 0.856 \\
ATE((1,0),(0,0)) & 0.501   & 0.409   & 0.698 & 0.801 & 0.761 \\
ATE((0,1),(0,0)) & 0.433  & 0.400   & 0.721 & 0.956  & 0.794 \\
ATE((1,0),(0,1)) & -0.626  & -0.420    & 0.731 & 0.759  & 0.852 \\ \hline
\end{tabular}\\
\scriptsize \noindent \textit{Note:} Simulations based on 10,000 Monte Carlo experiments. “Av. Bias”, “Med. Bias”, “RMSE”, “Cover” and “CIL’, stand for the average simulated bias, median simulated bias, simulated
root mean-squared errors, 95\% coverage probability, and 95\% confidence interval length,
respectively.
\end{table}

Table 4 shows the results to Monte Carlo simulations using the dynamic DiD estimator from De Chaisemartin and D'Haultefoueille (2023). Once again estimates are severely biased, which is also expected as strong exogeneity and parallel trends are violated in this setting. Confidence intervals are the largest among the considered procedures, which explains why  mean coverage is closer to 95\% than mean coverage using the g-formula.

\section{Conclusion}
In this paper I propose a novel identification method for marginal treatment effects and average treatment effects in models for dynamic treatment sequences and outcomes. I show that under a dynamic version of the standard exclusion restriction treatment effects are identified by combining the intuition from the cross-sectional MTE framework with the g-formula method that accounts for time-varying confounders. The identification strategy relaxes the two-way exclusion restriction and the sequential rank similarity assumptions previously required for point identification in these settings.

Monte Carlo simulation studies assert the desirable finite-sample properties of a series estimator for the dynamic ATEs when applied to a DGP tailored to resemble a well known empirical application. The estimator for the dynamic ATEs also exhibits good finite-sample performance, while alternative estimators in the sequential treatments literature fail to recover the true parameters.

This is a very preliminary draft for the first steps of this research project. A natural direction for next steps is to proceed with the identification of different treatment effects in settings with sequential treatments, such as dynamic versions of the local average treatment effect (LATE), the average treatment effect on the treated and on the untreated (ATT and ATU), and the policy relevant treatment effect (PRTE).

Another important direction to advance in the near future is to advance in the understanding of how treatment effects vary with time. The structural model presented is flexible enough to allow for heterogeneity associated with the timing of treatment, but it is still not clear how to disentangle the time-to-treatment effects of a particular choice treatment from the effects of the full sequence of treatment. Advancing in this question would allow for the possibility of identification of "event-study" parameters.

The Monte Carlo simulations suggest consistency of the estimator, but the formal derivation of the asymptotic properties of the estimator for the treatment effects is necessary and already in progress.

Finally, an empirical application for the procedure is also on the pipeline. I requested the data from the Fast Track Program, but is not made available until the time of submission of this draft. Other applications are also being taken into consideration. For example, there is data available for some of the papers on charter school effects that could be used to show the empirical appeal of this identification strategy.

%Bibliography
\bibliographystyle{apalike}  
\bibliography{templateArxiv}  

\section*{Appendix}

\subsection*{No-identification using a dynamic LIV estimand}

Once again, covariates are suppressed from notation but results hold conditional on X. I begin by writing the realized outcome in terms of potential outcomes:

\begin{align*}
    \mathbb{E}\left [ Y_{2}|Y_{1},\pi_{1}(Z_{1}),\pi_{2}(\textbf{Z}^{2},Y_{1},D_{1}) \right ]\\
=\mathbb{E}\left [ Y_{2}(0,0)|Y_{1},\pi_{1}(Z_{1}),\pi_{2}(\textbf{Z}^{2},Y_{1},D_{1}) \right ]\\+\mathbb{E}\left [D_{1}D_{2} (Y_{2}(1,1)+Y_{2}(0,0)-Y_{2}(0,1)-Y_{2}(1,0))|Y_{1},\pi_{1}(Z_{1}),\pi_{2}
(\textbf{Z}^{2},Y_{1},D_{1}) \right ]\\
+\mathbb{E}\left [D_{2} (Y_{2}(0,1)-Y_{2}(0,0))|Y_{1},\pi_{1}(Z_{1}),\pi_{2}
(\textbf{Z}^{2},Y_{1},D_{1}) \right ]\\
+\mathbb{E}\left [D_{1} (Y_{2}(1,0)-Y_{2}(0,0))|Y_{1},\pi_{1}(Z_{1}),\pi_{2}
(\textbf{Z}^{2},Y_{1},D_{1}) \right ]
\end{align*}

Under Assumption $\textbf{SX}$ this conditional expectation is equal to

\begin{align*}
    \mathbb{E}\left [ Y_{2}(0,0)|Y_{1}\right ]\\+\mathbb{E}\left [ Y_{2}(1,1)+Y_{2}(0,0)-Y_{2}(0,1)-Y_{2}(1,0)|Y_{1},D_{1}=1,\pi_{1}(Z_{1}),D_{2}=1,\pi_{2}
(\textbf{Z}^{2},Y_{1},D_{1}) \right ]\pi_{1}(Z_1)\pi_{2}
(\textbf{Z}^{2},Y_{1},D_{1})\\
+\mathbb{E}\left [Y_{2}(0,1)-Y_{2}(0,0)|Y_{1},\pi_{1}(Z_{1}),D_{2}=1,\pi_{2}
(\textbf{Z}^{2},Y_{1},D_{1}) \right ]\pi_{2}
(\textbf{Z}^{2},Y_{1},D_{1})\\
+\mathbb{E}\left [ (Y_{2}(1,0)-Y_{2}(0,0)|Y_{1},D_{1}=1,\pi_{1}(Z_{1}),\pi_{2}
(\textbf{Z}^{2},Y_{1},D_{1}) \right ]\pi_{1}(Z_{1})
\end{align*}

Which is equal to

\begin{align*}
    \mathbb{E}\left [ Y_{2}(0,0)|Y_{1}\right ]\\+\mathbb{E}\left [ Y_{2}(1,1)+Y_{2}(0,0)-Y_{2}(0,1)-Y_{2}(1,0)|Y_{1},V_{1}\leq\pi_{1}(Z_{1}),V_{2}\leq\pi_{2}
(\textbf{Z}^{2},Y_{1},D_{1}) \right ]\pi_{1}(Z_1)\pi_{2}
(\textbf{Z}^{2},Y_{1},D_{1})\\
+\mathbb{E}\left [Y_{2}(0,1)-Y_{2}(0,0)|Y_{1},\pi_{1}(Z_{1}),V_{2}\leq\pi_{2}
(\textbf{Z}^{2},Y_{1},D_{1}) \right ]\pi_{2}
(\textbf{Z}^{2},Y_{1},D_{1})\\
+\mathbb{E}\left [ (Y_{2}(1,0)-Y_{2}(0,0)|Y_{1},V_{1}\leq\pi_{1}(Z_{1}),\pi_{2}
(\textbf{Z}^{2},Y_{1},D_{1}) \right ]\pi_{1}(Z_{1})
\end{align*}

which equals to

\begin{align*}
    \mathbb{E}\left [ Y_{2}(0,0)|Y_{1}\right ]\\+\int_{0}^{\pi_{1}(Z_{1})}\int_{0}^{\pi_{2}(\textbf{Z}^{2},Y_{1},D_{1})}\mathbb{E}\left [ Y_{2}(1,1)+Y_{2}(0,0)-Y_{2}(0,1)-Y_{2}(1,0)|Y_{1},V_{1}=v_{1},V_{2}=v_{2} \right ]dv_{1}dv_{2}
\\
+\int_{0}^{\pi_{2}(\textbf{Z}^{2},Y_{1},D_{1})}\mathbb{E}\left [Y_{2}(0,1)-Y_{2}(0,0)|Y_{1},\pi_{1}(Z_{1}),V_{2}=v_{2} \right ]dv_{2}\\
+\int_{0}^{\pi_{1}(Z_{1})}\mathbb{E}\left [ (Y_{2}(1,0)-Y_{2}(0,0)|Y_{1},V_{1}=v_{1},\pi_{2}
(\textbf{Z}^{2},Y_{1},D_{1}) \right ]dv_{1}
\end{align*}

The result follows from the Leibniz integral rule

\subsection*{Proof of Theorem 1}

We suppress covariates from the exposition, but results hold conditional on $X=x$. First, note that from Assumption \textbf{NA},

\begin{align*}
    \mathbb{E}\left [ \mathbf{1}\left \{ \textbf{D}^{t}=\textbf{d}^{t} \right \}Y_{t}|\textbf{Y}^{t-1},\pi_{1}(Z_{1}),...,\pi_{t}(\textbf{Y}^{t-1},\textbf{D}^{t-1},\textbf{Z}^{t}) \right ]\\=\mathbb{E}\left [ \mathbf{1}\left \{ \textbf{D}^{t}=\textbf{d}^{t} \right \}Y_{t}(\textbf{d}^{t})|\textbf{Y}^{t-1},\pi_{1}(Z_{1}),...,\pi_{t}(\textbf{Y}^{t-1},\textbf{D}^{t-1},\textbf{Z}^{t}) \right ]\\=\mathbb{E}\left [ Y_{t}(\textbf{d}^{t})|\textbf{Y}^{t-1},\textbf{D}^{t}=\textbf{d}^{t},\pi_{1}(Z_{1}),...,\pi_{t}(\textbf{Y}^{t-1},\textbf{D}^{t-1},\textbf{Z}^{t}) \right ]Pr(\textbf{D}^{t}=\textbf{d}^{t}|\pi_{1}(Z_{1}),...,\pi_{t}(\textbf{Y}^{t-1},\textbf{D}^{t-1},\textbf{Z}^{t}))
\end{align*}

First, we focus on the probability $Pr(\textbf{D}^{t}=\textbf{d}^{t}|\pi_{1}(Z_{1}),...,\pi_{t}(\textbf{Y}^{t-1},\textbf{D}^{t-1},\textbf{Z}^{t}))$. Note that under assumptions $\textbf{C}$ and $\textbf{SX}$, it can be decomposed as

\begin{equation*}
    Pr(\textbf{D}^{t}=\textbf{d}^{t}|\pi_{1}(Z_{1}),...,\pi_{t}(\textbf{Y}^{t-1},\textbf{D}^{t-1},\textbf{Z}^{t}))=Pr(D_{1}=d_{1}|\pi_{1}(Z_{1}))...Pr(D_{t}=d_{t}|\pi_{t}(\textbf{Y}^{t-1},\textbf{D}^{t-1},\textbf{Z}^{t})))
\end{equation*}

For any period $t=1,...,\mathcal{T}$,

\begin{equation*}
    Pr(D_{t}=d_{t}|\pi_{t}(\textbf{Y}^{t-1},\textbf{D}^{t-1},\textbf{Z}^{t}))=\left\{\begin{matrix}
\pi_{t}(\textbf{Y}^{t-1},\textbf{D}^{t-1},\textbf{Z}^{t}),\ if\ d_{t}=1 \\ 
1-\pi_{t}(\textbf{Y}^{t-1},\textbf{D}^{t-1},\textbf{Z}^{t}),\ if\ d_{t}=0
\end{matrix}\right.
\end{equation*}

Now, let's focus on the conditional expectation $\mathbb{E}\left [ Y_{t}(\textbf{d}^{t})|\textbf{Y}^{t-1},\textbf{D}^{t}=\textbf{d}^{t},\pi_{1}(Z_{1}),...,\pi_{t}(\textbf{Y}^{t-1},\textbf{D}^{t-1},\textbf{Z}^{t}) \right ]$. For some $t^{'}\leq t$, assume $d_{t^{'}}$=1. Let $\textbf{D}^{t}_{-t^{'}}$ denote the sequence of treatment status until period $t$ without period $t^{'}$ and $\pi^{t}_{-t^{'}}(.)$ denote the sequence of propensity scores until period $t$ without period $t^{'}$. Under Assumption $\textbf{SX}$, we can write

\begin{align*}
    \mathbb{E}\left [ Y_{t}(\textbf{d}^{t})|\textbf{Y}^{t-1},\textbf{D}^{t}=\textbf{d}^{t},\pi_{1}(Z_{1}),...,\pi_{t}(\textbf{Y}^{t-1},\textbf{D}^{t-1},\textbf{Z}^{t}) \right ]\\=\frac{1}{\pi_{t^{'}}(\textbf{Y}^{t^{'}-1},\textbf{D}^{t^{'}-1},\textbf{Z}^{t^{'}})}\int_{0}^{\pi_{t^{'}}(\textbf{Y}^{t^{'}-1},\textbf{D}^{t^{'}-1},\textbf{Z}^{t^{'}})}\mathbb{E}\left [ Y_{t}(\textbf{d}^{t})|\textbf{Y}^{t-1},\textbf{D}^{t}_{-t^{'}}=\textbf{d}^{t}_{-t^{'}},\pi^{t}_{-t^{'}}(.),V_{t^{'}}\leq\pi_{t^{'}}(\textbf{Y}^{t^{'}-1},\textbf{D}^{t^{'}-1},\textbf{Z}^{t^{'}}) \right ]dv_{t^{'}}
\end{align*}

Differentiating the integral with respect to $\pi_{t^{'}}(\textbf{Y}^{t^{'}-1},\textbf{D}^{t^{'}-1},\textbf{Z}^{t^{'}})$ yields

\begin{equation*}
    \mathbb{E}\left [ Y_{t}(\textbf{d}^{t})|\textbf{Y}^{t-1},\textbf{D}^{t}_{-t^{'}}=\textbf{d}^{t}_{-t^{'}},\pi^{t}_{-t^{'}}(.),V_{t^{'}}=\pi_{t^{'}}(\textbf{Y}^{t^{'}-1},\textbf{D}^{t^{'}-1},\textbf{Z}^{t^{'}}) \right ]
\end{equation*}

If $d_{t^{'}}=0$, then

\begin{align*}
    \mathbb{E}\left [ Y_{t}(\textbf{d}^{t})|\textbf{Y}^{t-1},\textbf{D}^{t}=\textbf{d}^{t},\pi_{1}(Z_{1}),...,\pi_{t}(\textbf{Y}^{t-1},\textbf{D}^{t-1},\textbf{Z}^{t}) \right ]\\=\frac{1}{1-\pi_{t^{'}}(\textbf{Y}^{t^{'}-1},\textbf{D}^{t^{'}-1},\textbf{Z}^{t^{'}})}\int_{\pi_{t^{'}}(\textbf{Y}^{t^{'}-1},\textbf{D}^{t^{'}-1},\textbf{Z}^{t^{'}})}^{1}\mathbb{E}\left [ Y_{t}(\textbf{d}^{t})|\textbf{Y}^{t-1},\textbf{D}^{t}_{-t^{'}}=\textbf{d}^{t}_{-t^{'}},\pi^{t}_{-t^{'}}(.),V_{t^{'}}>\pi_{t^{'}}(\textbf{Y}^{t^{'}-1},\textbf{D}^{t^{'}-1},\textbf{Z}^{t^{'}}) \right ]dv_{t^{'}}
\end{align*}

Differentiating the integral yields

\begin{equation*}
    -\mathbb{E}\left [ Y_{t}(\textbf{d}^{t})|\textbf{Y}^{t-1},\textbf{D}^{t}_{-t^{'}}=\textbf{d}^{t}_{-t^{'}},\pi^{t}_{-t^{'}}(.),V_{t^{'}}=\pi_{t^{'}}(\textbf{Y}^{t^{'}-1},\textbf{D}^{t^{'}-1},\textbf{Z}^{t^{'}}) \right ]
\end{equation*}

Therefore, differentiating with respect to the propensity scores at all periods until $t$ yields

\begin{equation*}
    \mathbb{E}\left [ Y_{t}(\textbf{d}^{t})|\textbf{Y}^{t-1},V_{1}=\pi_{1}(Z_{1}),...,V_{t}=\pi_{t}(\textbf{Y}^{t-1},\textbf{D}^{t-1},\textbf{Z}^{t})\right ]
\end{equation*}

if the number of untreated periods is even (the negative signs from the derivatives with respect to the propensity score from untreated periods cancel out), and 

\begin{equation*}
    \mathbb{E}\left [ Y_{t}(\textbf{d}^{t})|\textbf{Y}^{t-1},V_{1}=\pi_{1}(Z_{1}),...,V_{t}=\pi_{t}(\textbf{Y}^{t-1},\textbf{D}^{t-1},\textbf{Z}^{t})\right ]
\end{equation*}

if the number of untreated periods is odd.

The fractions from that multiply the integral cancel out with $Pr(\textbf{D}^{t}=\textbf{d}^{t}|\pi_{1}(Z_{1}),...,\pi_{t}(\textbf{Y}^{t-1},\textbf{D}^{t-1},\textbf{Z}^{t}))$, and thus it follows that

\begin{equation*}
    \frac{\partial \mathbb{E}\left [ \mathbf{1}\left \{ \textbf{D}^{t}=\textbf{d}^{t} \right \}Y_{t}|\textbf{Y}^{t-1},\pi^{t}(\textbf{Y}^{t-1},\textbf{D}^{t-1},\textbf{Z}^{t}) \right ]}{\partial \pi_{1}(Z_{1})...\partial\pi_{t}(\textbf{Y}^{t-1},\textbf{D}^{t-1},\textbf{Z}^{t})}=\mathbb{E}\left [ Y_{t}(\textbf{d}^{t})|\textbf{Y}^{t-1},V_{1}=\pi_{1}(Z_{1}),...,V_{t}=\pi_{t}(\textbf{Y}^{t-1},\textbf{D}^{t-1},\textbf{Z}^{t})\right ]
\end{equation*}

when the number of untreated periods is even and 

\begin{equation*}
    -\frac{\partial \mathbb{E}\left [ \mathbf{1}\left \{ \textbf{D}^{t}=\textbf{d}^{t} \right \}Y_{t}|\textbf{Y}^{t-1},\pi^{t}(\textbf{Y}^{t-1},\textbf{D}^{t-1},\textbf{Z}^{t}) \right ]}{\partial \pi_{1}(Z_{1})...\partial\pi_{t}(\textbf{Y}^{t-1},\textbf{D}^{t-1},\textbf{Z}^{t})}=\mathbb{E}\left [ Y_{t}(\textbf{d}^{t})|\textbf{Y}^{t-1},V_{1}=\pi_{1}(Z_{1}),...,V_{t}=\pi_{t}(\textbf{Y}^{t-1},\textbf{D}^{t-1},\textbf{Z}^{t})\right ]
\end{equation*}

when the number of untreated periods is odd.

Under Assumption $\textbf{SX}$, the CDF 

\begin{equation*}
    F_{Y_{t}|\textbf{Y}^{t-1},\textbf{D}^{t-1},\textbf{Z}^{t}}(y_{t}|\textbf{y}^{t-1},\textbf{d}^{t-1},\textbf{z}^{t})=F_{Y_{t}|\textbf{Y}^{t-1},\textbf{D}^{t-1},\pi^{t}(\textbf{Y}^{t-1},\textbf{D}^{t-1},\textbf{Z}^{t})}(y_{t}|\textbf{y}^{t-1},\textbf{d}^{t-1},\pi^{t}(\textbf{y}^{t-1},\textbf{d}^{t-1},\textbf{z}^{t}))
\end{equation*}

is identified for any period $t=1,...,\mathcal{T}$.

Thus, it follows that

\begin{align*}
    \int_{\textbf{y}^{t-1}}\frac{\partial \mathbb{E}\left [ \mathbf{1}\left \{ \textbf{D}^{t}=\textbf{d}^{t} \right \}Y_{t}|\textbf{Y}^{t-1},\pi^{t}(\textbf{Y}^{t-1},\textbf{D}^{t-1},\textbf{Z}^{t}) \right ]}{\partial \pi_{1}(Z_{1})...\partial\pi_{t}(\textbf{Y}^{t-1},\textbf{D}^{t-1},\textbf{Z}^{t})}\prod_{k=1}^{t-1}dF_{Y_{k}|\textbf{Y}^{k-1},\textbf{D}^{k-1},\textbf{Z}^{k}}(y_{k}|\textbf{y}^{k-1},\textbf{d}^{k-1},\textbf{d}^{k})\\=\mathbb{E}\left [ Y_{t}|V_{1}=\pi_{1}(Z_{1}),...,V_{t}=\pi_{t}(\textbf{Y}^{t-1},\textbf{D}^{t-1},\textbf{Z}^{t}) \right ]=m_{\textbf{d}^{t}}(\pi^{t}(\textbf{Y}^{t-1},\textbf{D}^{t-1},\textbf{Z}^{t}))
\end{align*}

for and even number of untreated periods and

\begin{align*}
    \int_{\textbf{y}^{t-1}}-\frac{\partial \mathbb{E}\left [ \mathbf{1}\left \{ \textbf{D}^{t}=\textbf{d}^{t} \right \}Y_{t}|\textbf{Y}^{t-1},\pi^{t}(\textbf{Y}^{t-1},\textbf{D}^{t-1},\textbf{Z}^{t}) \right ]}{\partial \pi_{1}(Z_{1})...\partial\pi_{t}(\textbf{Y}^{t-1},\textbf{D}^{t-1},\textbf{Z}^{t})}\prod_{k=1}^{t-1}dF_{Y_{k}|\textbf{Y}^{k-1},\textbf{D}^{k-1},\textbf{Z}^{k}}(y_{k}|\textbf{y}^{k-1},\textbf{d}^{k-1},\textbf{d}^{k})\\=\mathbb{E}\left [ Y_{t}|V_{1}=\pi_{1}(Z_{1}),...,V_{t}=\pi_{t}(\textbf{Y}^{t-1},\textbf{D}^{t-1},\textbf{Z}^{t}) \right ]=m_{\textbf{d}^{t}}(\pi^{t}(\textbf{Y}^{t-1},\textbf{D}^{t-1},\textbf{Z}^{t}))
\end{align*}

for an odd number of untreated periods, which concludes the proof.

\subsection*{Details of the Monte Carlo simulation}

Here, I show details of the Data Generating Process (DGP) used in the Monte Carlo simulations. I simulate panel data with two periods ($\mathcal{T}=2$), and sample size $n$ at each period equal to 891, which is the number of children that were part of the actual Fast Track Program.

I simulate two binary covariates ($X_{1}, X_{2}$) using Bernoulli distribtuions with parameters 0.5 and 0.6 respectively. The continuous covariate ($X_{3})$ is generated through a normal distribution with mean equal to 5 and variance equal to 1.2. The time-specific instruments ($Z_{1},Z_{2})$ were generated from independent standard normal distributions.

The propensity score for each period are logistic:

\begin{align*}
    \pi_{1}(Z_{1},X)=\frac{exp(Z_{1}+\alpha_{1}^{'}X)}{1+exp(Z_{1}+\alpha_{1}^{'}X)}\\\pi_{1}(Z^{2},D_{1},Y_{1},X)=\frac{exp(Z_{1}+Z_{2}+\alpha_{2}^{'}X+D_{1}-Y_{1})}{1+exp(Z_{1}+Z_{2}+\alpha_{2}^{'}X+D_{1}-Y_{1})}
\end{align*}

where $\alpha_{1}=\alpha_{2}=(-0.25,-0.15,0.4)$.

The error terms from the choice models ($V_{1},V_{2})$ are generated from independent uniform distributions. Selection into treatment at each period follows 

\begin{align*}
    D_{1}=\mathbf{1}\left \{ \pi_{1}(Z_{1},X)\geq V_{1} \right \}\\D_{2}=\mathbf{1}\left \{ \pi_{1}(\textbf{Z}^{2},Y_{1},D_{1},X)\geq V_{2} \right \}
\end{align*}

The error terms from potential outcomes are normally distributed and linearly correlated to the error terms from the choice models. In period 1, I specify $U_{1}(1)\sim\mathcal{N}(2(V_{1}-0.5),1)$ and $U_{1}(0)\sim\mathcal{N}(V_{1}-0.5,1)$. For the second period, I specify $U_{2}(1,d_{1})\sim\mathcal{N}(V_{2}-0.5+U_{1}(d_{1}),1)$ and $U_{2}(0,d_{1})\sim\mathcal{N}(2(V_{2}-0.5)+U_{1}(d_{1}),1)$.

Potential outcomes binary and covariates enter linearly. For the first period:

\begin{align*}
    Y_{1}(1)=\mathbf{1}\left \{ \beta_{1}^{(1)'}X\geq U_{1}(1) \right \}\\Y_{1}(0)=\mathbf{1}\left \{ \beta_{1}^{(0)'}X\geq U_{1}(0) \right \}
\end{align*}

In the second period, the previous outcome also enters linearly in the working model:

\begin{align*}
    Y_{2}(1,d_{1})=\mathbf{1}\left \{ \beta_{2}^{(1,d_{1})'}X+0.1Y_{1}\geq U_{2}(1,d_{1}) \right \}\\Y_{2}(0,d_{1})=\mathbf{1}\left \{ \beta_{2}^{(0,d_{1})'}X+0.1Y_{1}\geq U_{2}(0,d_{1}) \right \}
\end{align*}

where $\beta_{1}^{(1)}=\beta_{1}^{(0)}=\beta_{2}^{(1,d_{1})}=\beta_{2}^{(1,d_{1})}=(-1.2,-1.5,0.4)$.

Estimation of marginal treatment responses combines a series estimator for the marginal treatment responses conditional on $Y_{1}$ and a two-stage least squares estimator for $Y_{1}$. The estimators are put in a plug-in estimator for the g-formula for MTRs.

For the conditional expectiation, we have

\begin{align*}
    \mathbb{E}\left [ \mathbf{1}\left \{ \textbf{D}^{2}=\textbf{d}^{2} \right \}Y_{2}|X,Y_{1},\pi_{1}(Z_{1},X),\pi_{2}(\textbf{Z}^{2},Y_{1},D_{1},X) \right ]\\=Pr(\textbf{D}^{2}=\textbf{d}^{2}|X,\pi_{1}(Z_{1},X),\pi_{2}(\textbf{Z}^{2},Y_{1},D_{1},X))\left ( \beta_{2}^{\textbf{d}^{2}}X+\theta Y_{1}+\mathbb{E}\left [ U_{2}(\textbf{d}^{2})| \right ] \pi_{1}(Z_{1},X),\pi_{2}(\textbf{Z}^{2},Y_{1},D_{1},X)\right )\\\approx Pr(\textbf{D}^{2}=\textbf{d}^{2}|X,\pi_{1}(Z_{1},X),\pi_{2}(\textbf{Z}^{2},Y_{1},D_{1},X))\left ( \beta_{2}^{\textbf{d}^{2}}X+\theta Y_{1}+P_{k}(\pi_{1}(Z_{1},X),\pi_{2}(\textbf{Z}^{2},Y_{1},D_{1},X))\right )
\end{align*}

Where $P_{k}$ is a vector of approximating functions for the error term.

I approximate the MTR conditional on $Y_{1}$ by

\begin{align*}
    \mathbb{E}\left [Y_{2}|X,Y_{1},V_{1}=\pi_{1}(Z_{1},X),V_{2}=\pi_{2}(\textbf{Z}^{2},Y_{1},D_{1},X) \right ]\\\approx \triangledown_{\pi_{1},\pi_{2}}\mathbb{E}\left [ \mathbf{1}\left \{ \textbf{D}^{2}=\textbf{d}^{2} \right \}Y_{2}|X,Y_{1},\pi_{1}(Z_{1},X),\pi_{2}(\textbf{Z}^{2},Y_{1},D_{1},X) \right ]
\end{align*}

where $\triangledown_{\pi_{1},\pi_{2}}$ denotes the gradient of the approximating functions with respect to $\pi_{1}$ and $\pi_{2}$. I then combine the estimates from the series estimator with the two-stage least squares estimator to obtain the semiparametric g-formula estimate for the MTRs.

\end{document}